%% file: article_v13_arXiv.tex
\documentclass[reprint,amsmath,amssymb,aps,prl,longbibliography]{revtex4-2}

\usepackage[utf8]{inputenc} \usepackage[T1]{fontenc}
\usepackage[english]{babel} \usepackage[autostyle=true]{csquotes}
\usepackage[colorlinks=true,  urlcolor=blue,  linkcolor=blue,citecolor=blue]{hyperref}
\usepackage[dvipsnames]{xcolor} \usepackage{siunitx} \usepackage{braket} 
\usepackage{graphicx}
\usepackage{bm}
\usepackage{todonotes}
\usepackage{physics}
\usepackage{txfonts}

\definecolor{red1}{RGB}{175,0,0}
\definecolor{violet1}{RGB}{150,0,75}
\definecolor{green1}{RGB}{0,150,0}
\definecolor{blue1}{HTML}{1A237E}

\DeclareMathOperator{\e}{e}
\newcommand{\mi}{\mathrm{i}}
\newcommand\bnabla{\boldsymbol{\nabla}}
\newcommand{\mum}{\rm \si{\micro\meter}}

\begin{document}
    
\title{Superfluid Fraction of a 2D Bose-Einstein Condensate in a Triangular Lattice}

\author{F. Rabec, G. Brochier, S. Wattellier, G. Chauveau, Y. Li, S. Nascimbene, J. Dalibard, J. Beugnon}

\email{beugnon@lkb.ens.fr}

\affiliation{Laboratoire Kastler Brossel,  Coll\`ege de France, CNRS, ENS-PSL
University, Sorbonne Universit\'e, 11 Place Marcelin Berthelot, 75005 Paris,
France}

\date{\today}
\begin{abstract}
We experimentally investigate the superfluid properties of a two-dimensional, weakly interacting
Bose-Einstein condensate in the zero-temperature regime, when it is subjected to a triangular
optical lattice potential. We implement an original method, which involves solving the hydrodynamic
continuity equation to extract the superfluid fraction tensor from the measured \textit{in situ}
density distribution of the fluid at rest. In parallel, we apply an independent dynamical approach
that combines compressibility and sound velocity measurements to determine the superfluid fraction.
Both methods yield consistent results in good agreement with simulations of the
Gross-Pitaevskii equation as well as with the Leggett bounds determined from  the measured density profiles.
\end{abstract}

\maketitle
\input{main_v18.tex}
\definecolor{violet3}{RGB}{70,0,120}
\hypersetup{urlcolor=violet3} 

\bibliography{bibliography}
\onecolumngrid
\newpage
\section{End Matter}
\twocolumngrid
\input{End_matter_v9.tex}
\clearpage
\section{Supplemental material}
\input{sm_v13.tex}

\end{document}

%% file: main_v18.tex
\paragraph{\textcolor{red1}{Introduction--}}
Superfluid states of matter have been studied in many different settings, including liquid helium, atomic gases and photonic systems\,\cite{Kapitza38,Allen38,Baym69,Raman99,Zwierlein05,Amo09,Fontaine20}. They are usually characterized by their response to a mobile external perturbation, such as the motion of the container holding them\,\cite{Leggett99}. This response can be decomposed into normal and superfluid contributions. The weight of the superfluid component is defined as the superfluid fraction $f_s$, a quantity which was first introduced in the context of finite-temperature systems\,\cite{pitaevskiiBoseEinstein2016}. It has been measured in liquid helium (\textit{e.g.,} in Refs.\,\cite{Greywall73,Maynard1976}) and more recently in atomic gases \cite{Sidorenkov13,Christodoulou21,Hilker22,Yan24}, by studying the propagation of second sound.

At zero temperature, Galilean-invariant superfluids exhibit a superfluid fraction of unity, whereas in spatially modulated systems the superfluid fraction is reduced. Density modulations can be imposed by an external potential or occur spontaneously, as in supersolids. Determining $f_s$ in these systems has attracted increasing interest since the realization of supersolid states in atomic gases \,\cite{Li17,Leonard17,Bottcher19,Chomaz19,Tanzi19} and polariton condensates \,\cite{Trypogeorgos25}. Recent experiments have reported the measurement of $f_s$ in a 2D Bose gas modulated along one direction\,\cite{chauveauSuperfluid2023,taoObservation2023}, in a 1D supersolid\,\cite{Tanzi21,Biagioni24}, in a molecular BEC\,\cite{Pezze24} and in a driven superfluid\,\cite{Liebster25}. The superfluid fraction of zero-temperature disordered Bose gases\,\cite{geierSuperfluidity2024,perez-cruzSuperfluid2024}, strongly interacting Fermi gases\,\cite{Orso24}, and neutron stars\,\cite{Chamel24,Almirante24} has also been studied theoretically.

In a two-dimensional modulated system, superfluidity is characterized by a superfluid tensor. It is defined from the average momentum $\langle\hat{\boldsymbol{P}}\rangle$ of a system subjected to the perturbation $-\boldsymbol{v}_0 \cdot \hat{\boldsymbol{P}}$, corresponding to the motion of an external potential with velocity $\boldsymbol{v}_0$: 
\begin{equation}
    f_{s}^{(i,j)} =  \delta_{ij} - \lim_{\boldsymbol{v}_0 \to 0} \frac{\langle\hat{P}_j\rangle}{N M v_{0,i}}, \label{eq:fs_definition}
\end{equation} 
where $N$ is the number of  particles, $M$ is the atomic mass, and the indices $i,j$ are associated with the Cartesian coordinates\,\cite{blakieSuperfluid2024,geierSuperfluidity2024,perez-cruzSuperfluid2024}. 
In the experiments reported in Refs.\,\cite{chauveauSuperfluid2023,taoObservation2023}, which studied a 2D gas modulated along the $x$ direction, this tensor is diagonal in the $(x,y)$ basis, $f_{s}^{(y,y)}=1$ and $f_{s}^{(x,x)}$ is called the superfluid fraction. In these works, the superfluid fraction has been determined using various methods based on the measurement of the speed of sound, the study of scissor modes, and the calculation of the Leggett bounds from the density profile\,\cite{leggettCan1970,leggettSuperfluid1998}. Such methods are well suited to 1D modulation or to cases where the density profile is a separable function of $x$ and $y$. However, these methods are not generally applicable to a 2D modulation. Therefore,  measuring the superfluid tensor in 2D modulated systems remains an open problem, particularly in light of  the recent realization of 2D supersolids\,\cite{Norcia21}.

\begin{figure}[t!!!]
    \begin{center}
\vskip10pt
   \includegraphics[trim={0cm 0cm 0 0cm},clip,width=\columnwidth]{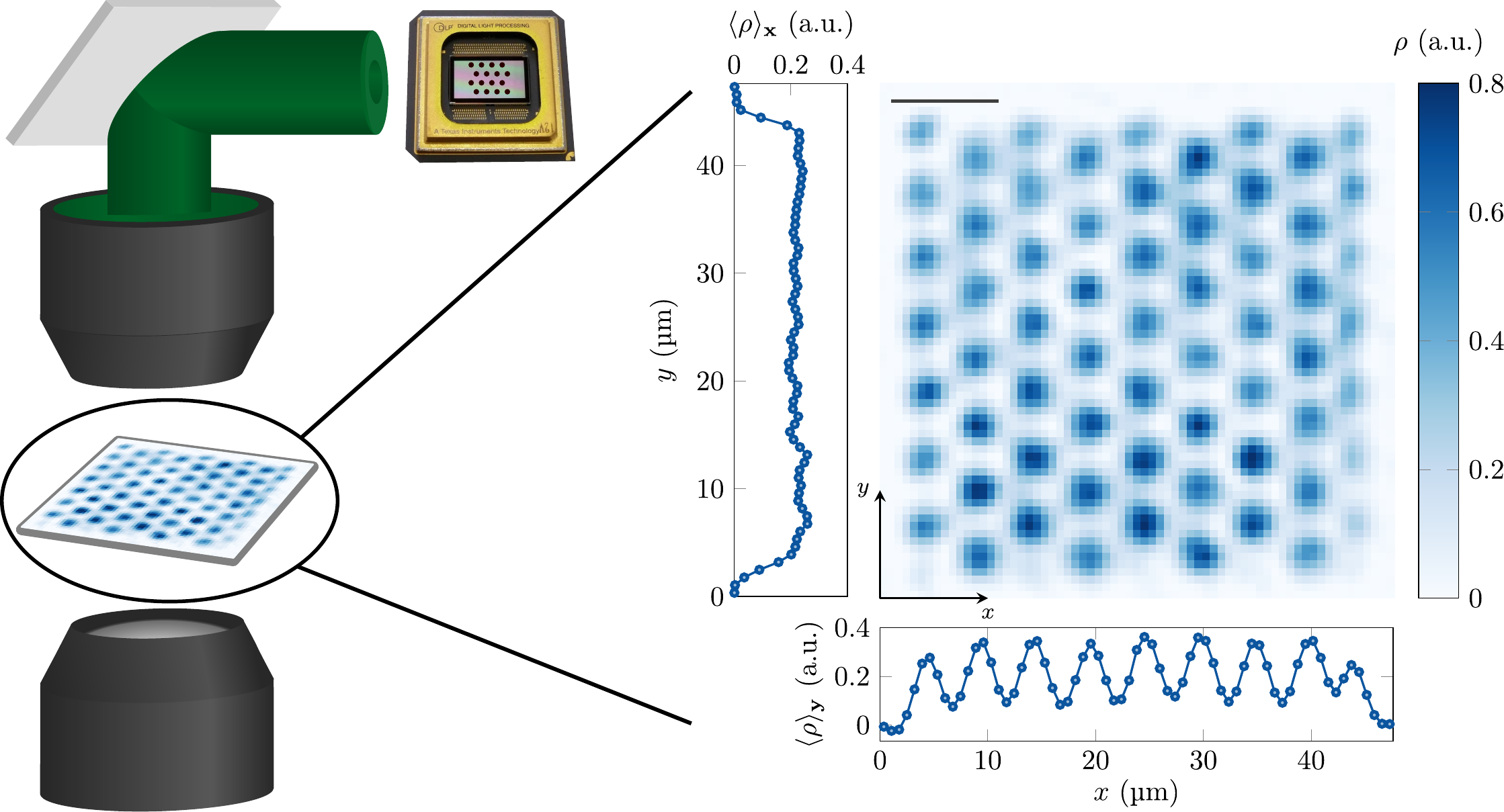} 
    \end{center}
    \caption{\textbf{Sketch of the experiment and density distribution.} A triangular lattice potential is projected onto a 2D Bose-Einstein condensate using  a spatial light modulator and repulsive light. High-resolution imaging is used to determine the atomic density distribution. The zoomed picture is an absorption image, averaged 40 times, of the atomic cloud subjected to a lattice potential with a depth $V_0\simeq4.7\,\mu_0$, where $\mu_0/k_B = 45(2)\,$nK is the chemical potential of the cloud in the absence of the lattice potential. The color bar encodes the surface density. The length of the scale bar is $\SI{10}{\um}$. Integrated density profiles along the $x$ and $y$ directions are also shown.}\label{fig:lattice}
\end{figure}

In this work, we apply a triangular lattice potential to a 2D, low-temperature, weakly-interacting Bose condensed gas and we study its superfluid fraction as a function of the lattice amplitude. We first introduce  the Leggett bounds and compute them from the density profile of the modulated system at rest. We then describe a method, first introduced by W.\,M.~Saslow\,\cite{saslowSuperfluidity1976}, that is based on solving the continuity equation for the fluid, to determine the superfluid fraction tensor solely from the density profile. Finally, we present a method based on a dynamic measurement that combines the determination of the cloud compressibility and the speed of sound in order to compute the superfluid fraction. These two approaches yield consistent results and we provide a quantitative comparison with numerical simulations of the Gross-Pitaevskii Equation (GPE), which accurately describes our weakly-interacting BEC.

\vskip5pt
\paragraph{\textcolor{red1}{Experimental system--} }
The experimental platform has already been described in Refs.~\cite{villeLoading2017, villeSound2018,zouOptical2021, chauveauSuperfluid2023}. In brief, we first produce a $^{87}$Rb BEC in the magnetic sensitive hyperfine state $\ket{F=1,m_{F}=-1}$ of the electronic ground level. A quasi-2D Bose gas is then obtained by loading the cloud into a single node of a vertical optical lattice. This lattice provides a strong harmonic confinement along the vertical direction $z$ with a frequency of $\omega_z / 2 \pi  = \SI{4.1 \pm 0.2}{\kilo\hertz}$. The associated size of the ground-state wave function is $\ell_z = \sqrt{\hbar / M\omega_z} = \SI{168 \pm 4}{\nm}$. Introducing the \textit{s}-wave scattering length $a_s$, we obtain the 2D coupling constant $\tilde g=Mg/\hbar^2 =\sqrt{8\pi} a_s / \ell_z = 0.158(4)$ \cite{Bloch08}. An additional potential generated using a Digital Micromirror Device (DMD) creates a flat-bottom trap that confines our atoms to a square box of size $L=\SI{42\pm1}{\um}$. The average 2D density is fixed to $\rho_0=\SI{51 \pm 2}{\um^{-2}}$, which corresponds to a chemical potential $\mu_0/k_B=\SI{45\pm2}{nK}$. The temperature of the sample is below the lowest measurable value in our experiment, i.e., $< \SI{20}{\nano \kelvin}$, and the behavior of the gas is well approximated by the zero-temperature limit. We spatially modulate the density of the cloud in the $xy$ plane using an optical lattice generated using a second DMD. All trapping beams consist of blue-detuned repulsive light at wavelength $\lambda = \SI{532}{\nano \meter}$. The potential profile corresponds to a triangular lattice modulation given by
\begin{equation}
    V(\boldsymbol{r}) =\bar V - A_L \sum_{m=0}^{2} \cos(\boldsymbol{k}_m \cdot \boldsymbol{r}+\phi_m) 
\label{eq:perfect_triangular}
\end{equation}
with $\boldsymbol{k}_m = 4\pi/(\sqrt{3}d) \left\{ \cos\left(2m\pi/3\right), \sin\left(2m\pi/3\right) \right\}$, $\boldsymbol{r}=(x,y)$, $d= \SI{6 +- 0.1}{\micro \meter}$ the lattice period, $A_L$ the amplitude of each standing wave, and a global constant offset potential $\bar V$, which does not contribute to the studied dynamics \cite{Supplementary}. This potential has a three-fold symmetry. The global phase $\Phi=\phi_0+\phi_1+\phi_2$ characterizes the potential $V(\boldsymbol{r})$ of the unit cell of the lattice, whereas a variation of the $\phi_i$'s at constant $\Phi$ corresponds to a mere translation of the modulation pattern\,\footnote{The case $\Phi=0$ corresponds to the so-called ``triangular'' lattice potential (associated with the hexagonal lattice in Bravais classification), which has six-fold symmetry, and the case $\Phi=\pi$ to the honeycomb lattice potential.}. The lattice studied in this work is approximately 7 periods long and corresponds to $\Phi = \num{0.21 +- 0.01}\,\pi$. The peak-to-peak amplitude of the lattice modulation is denoted $V_0$ and we have $V_0 \simeq 5.0\,A_L$ for the chosen value of $\Phi$. An example of the obtained density distribution is shown in Fig.\,\ref{fig:lattice}. The projected potential is non-separable, but due to its three-fold symmetry, the superfluid tensor is expected to be isotropic (see End Matter). In the following, we will thus   restrict ourselves to measuring the superfluid fraction $f_s=f_{s}^{(y,y)}=f_{s}^{(x,x)}$ with $f_{s}^{(x,y)}=0$.


\begin{figure}[t!!]
    \begin{center}
                \includegraphics[width=0.95\columnwidth]{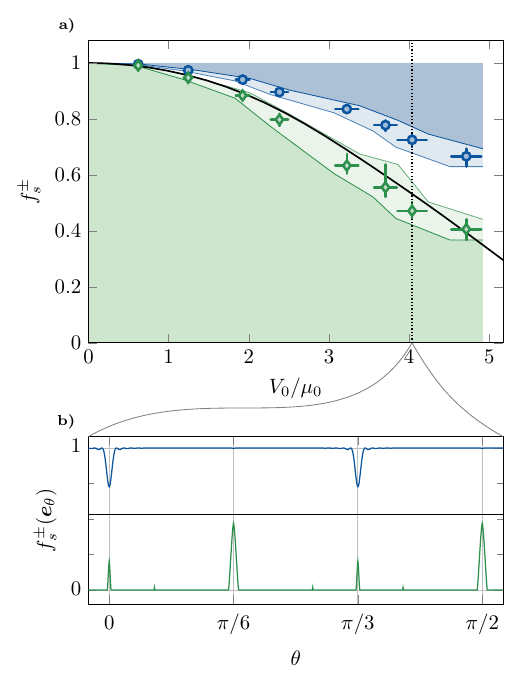}
    \end{center}
    \caption{
        \textbf{Leggett bounds.} 
        (a) Measured upper bound $f_s^{+} (\boldsymbol{e}_x)$ (blue circles) and lower bound $f_s^{-} (\boldsymbol{e}_y)$ (green diamonds).  Error bars correspond to the error propagation of the 1-$\sigma$ uncertainties in the calibration of the imaging system response.  The black solid line is the predicted superfluid fraction assuming a pure BEC described by the GPE. The colored regions define the excluded regions for the superfluid fraction according to Leggett bounds measurements. The darker-colored regions correspond to an exclusion beyond experimental errors. The white region shows the corresponding allowed region. The lighter-colored regions correspond to the intermediate region defined by the extension of experimental errors.
        (b) Experimentally determined upper $f_s^{+}(\boldsymbol{e}_{1})$ (solid blue) and lower $f_s^{-}(\boldsymbol{e}_{1})$ (solid green) Leggett bounds as a function of $\theta$, the angle between unit vector $\boldsymbol{e}_{1}$ and $\boldsymbol{e}_x$  for $V_0\simeq 4.0\,\mu_0$. The solid black line gives the value of $f_s$ already shown in (a) for this lattice depth. }
      \label{fig:Leggett_bounds} 
\end{figure}

\vskip5pt
\paragraph{\textcolor{red1}{Analysis of the density profiles--}}
Our study of the superfluid fraction requires an accurate determination of the \textit{in situ} density profile of the cloud. We use here absorption imaging. Due to the finite resolution of our imaging system, the measured density $\rho^{\text{(meas)}}(\boldsymbol{r})$ does not coincide with the actual density $\rho(\boldsymbol{r})$. We decompose the atomic density profile as $\rho(\boldsymbol{r})= \rho_0 + \sum_{n>0} \rho_n (\boldsymbol{r})$, where the index $n$ is associated with the Fourier components corresponding to a given  $\lambda_n$. We have $\lambda_1=d=6\,\mum$, $\lambda_2=d/\sqrt{3}\simeq3.46\,\mum$, etc. The imaging system acts as a low-pass filter, and we describe its contribution by a set of attenuation factors $\{ \beta_n\}_{n \in \mathbb N^\ast}$, which we calibrate independently up to $n=2$ (see \cite{Supplementary}). The measured density thus reads, $\rho^{\text{meas}}(\boldsymbol{r}) = \rho_0 + \sum_{n>0} \beta_n \rho_n (\boldsymbol{r})$. For the range of density modulation explored here, we found numerically that only the first two terms, of period $\lambda_1$ and $\lambda_2$, lead to a significant contribution. We fit the experimental data to this model for the first two terms of the sum, and then we reconstruct the actual density profile $\rho(\boldsymbol{r})$ of the cloud. The fitting function is given in Ref.\,\cite{Supplementary} and is chosen to ensure the isotropy of the superfluid tensor. 


\vskip5pt
\paragraph{\textcolor{red1}{Leggett bounds from the density profiles--}}
In his seminal works on supersolids, A.\,J.~Leggett introduced two quantities  $f_s^+$\cite{leggettCan1970} and $f_s^-$\,\cite{leggettSuperfluid1998}, which can be directly computed from the density profile and which provides upper and lower bounds to the superfluid fraction of the system, respectively. These bounds have different ranges of applicability. The upper bound, based on a variational approach, is valid for any fluid with time-reversal symmetry. The applicability range of the lower bound is more subtle. A sufficient condition is to apply it to a zero-temperature system for which a mean-field approach is valid, as is approximately the case for the weakly-interacting BEC studied here. For a gas modulated along a single direction, these bounds both coincide with the superfluid fraction, a property used in Ref.\,\cite{chauveauSuperfluid2023}. For a 2D modulation, we use
\begin{equation}
 f_s^{-} (\boldsymbol{e}_1) =    \dfrac{1}{\rho_0}\left\langle\frac{1}{\left\langle \dfrac{1}{\rho(\boldsymbol{r})}\right\rangle_{\hskip-3pt\boldsymbol{e}_1}}
\right\rangle{\raisebox{-15pt}{\scriptsize${\hskip-5pt\boldsymbol{e}_2}$}}\,\mathrm{and}\;
  f_s^{+} (\boldsymbol{e}_1)=\dfrac{1}{\rho_0} \frac{1}{\left\langle\dfrac{1}{\left\langle\rho(\boldsymbol{r})\right\rangle_{\hskip-1pt\boldsymbol{e}_2}}\right\rangle_{\hskip-3pt\boldsymbol{e}_1}},
    \label{eq:leggett_lower} 
\end{equation}
where $\left\langle\cdot\right\rangle_{\boldsymbol{e_i}}$ denotes the spatial average along direction ${\boldsymbol{e}}_i$ and the two directions $\boldsymbol{e}_1$ and $\boldsymbol{e}_2$ are taken to be orthogonal. For the two-dimensional system studied in this work, these bounds depend on the choice of the orientation of $\boldsymbol{e_1}(\theta)= \cos \theta \; \boldsymbol{e}_x + \sin \theta \; \boldsymbol{e}_y$, even in the case of the triangular lattice considered here for which the superfluid tensor is a scalar\,\cite{blakieSuperfluid2024}. We report in Fig.\,\ref{fig:Leggett_bounds}a the tightest bounds determined from the measured density profiles which, given the chosen orientation of the studied lattice (see Fig.\,\ref{fig:lattice}), are obtained for $f_s^{-} (\boldsymbol{e}_y)$  (\textit{i.e.}, $\theta=\pi/2$) and $f_s^{+}(\boldsymbol{e}_x)$ (\textit{i.e.}, $\theta=0$)\,\footnote{This is also the case for a triangular lattice with $\Phi=0$.}. The colored regions in Fig.\,\ref{fig:Leggett_bounds}a represent the values of the superfluid fraction that are excluded by these bounds. More precisely, the lighter-colored region is permitted within  the $1-\sigma$ experimental errors while the darker regions are excluded beyond them. The superfluid fraction predicted by the GPE is shown as a solid line and mostly lies in the white and lighter-colored regions, as expected. We also illustrate in Fig.\,\ref{fig:Leggett_bounds}b the angular dependence of Leggett bounds computed from the density profiles for $V_0\simeq 4.0\,\mu_0$~\cite{blakieSuperfluid2024}. This confirms the chosen angles to determine the tightest bounds (see \cite{Supplementary} for a comparison of the measured bounds to the GPE prediction). 

\begin{figure}[t!!]
    \begin{center}
                \includegraphics[width=0.95\columnwidth]{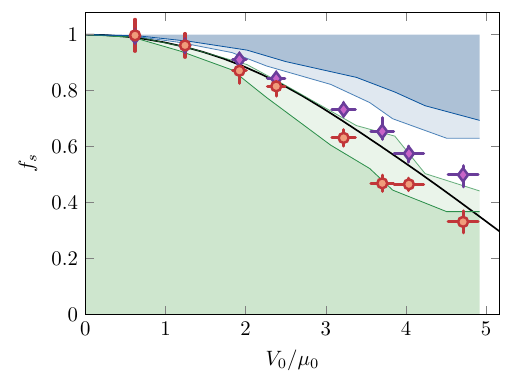}
    \end{center}
    \caption{\textbf{Superfluid fraction measurement.} Measured superfluid fraction as a function of the normalized lattice amplitude using the method based on the density profile (violet diamonds) and the dynamic approach (red circles).  Error bars for the density profile methods are determined as in Fig.\,\ref{fig:Leggett_bounds}. Error bars for the dynamic approach are deduced from the sound velocity and compressibility measurements. The solid line (same as in Fig.\,\ref{fig:Leggett_bounds}) is the predicted $f_s$. The shaded regions represent the excluded areas for the superfluid fraction according to the Leggett bounds measurements reported in Fig.\,\ref{fig:Leggett_bounds}.}\label{fig:superfluid_fraction} 
\end{figure}

\vskip5pt
\paragraph{\textcolor{red1}{Superfluid fraction  from the density profile--}}
We now turn to the determination of the superfluid fraction of the fluid directly from the measured density distribution\,\cite{saslowSuperfluidity1976,sepulvedaSuperfluid2010, blakieSuperfluid2024}. We introduce the static equilibrium many-body wave function $\psi_{\rm eq}(\boldsymbol{r})=\sqrt{\rho_{\rm eq}(\boldsymbol{r})}$ of the gas in the presence of the lattice potential. The superfluid fraction is associated with the modification of this wave function when switching to a reference frame moving at velocity $\boldsymbol{v}_0$. As the flow in this frame is time-independent, the continuity equation reads
\begin{equation}
\boldsymbol{\nabla}\cdot\left\{ \rho(\boldsymbol{r})\left[\boldsymbol{v}(\boldsymbol{r})-\boldsymbol{v}_0\right]\right\}=0,
\label{eq:continuity_equation}
\end{equation}
where the velocity field is associated with the phase profile, $S(\boldsymbol{r})$, of the wave function via $\boldsymbol{v}(\boldsymbol{r})= (\hbar/M) \boldsymbol{\nabla} S(\boldsymbol{r})$. At first order in Eq.\,\eqref{eq:continuity_equation}, one can replace $\rho(\boldsymbol r)$ by  $\rho_{\rm eq}(\boldsymbol r)$ and write the wave function as $\e^{\mi S(\boldsymbol{r})}\sqrt{\rho_{\rm eq}(\boldsymbol{r})}$. The numerical resolution of this continuity equation (see End Matter), with periodic boundary conditions, yields a single solution for the phase profile $S({\boldsymbol r})$. Using Eq.\,\eqref{eq:fs_definition}, one obtains the superfluid fraction tensor
\begin{equation}
    f_{s}^{(i,j)} = \delta_{ij} - \lim_{\boldsymbol{v}_0\to 0} \frac{\hbar}{N M v_{0,i}}\int \dd^2 r \, \rho_{\rm eq}(\boldsymbol{r})[\partial_j S(\boldsymbol{r})]. 
    \label{eq:fs_definition_phi}
\end{equation}
We apply this approach to our experimental data to determine the superfluid fraction tensor. We show in Fig.\,\ref{fig:superfluid_fraction} (diamonds)  $f_s^{(y,y)} \equiv f_s$,  which is in good agreement with the predicted one\,\footnote{By construction, for each experimental density profile, the superfluid fraction computed by this method lies between the two Leggett bounds. For the dynamic approach, as we use independent data, this constraint does not apply.}. We also performed an alternative analysis of the density profiles, allowing all Fourier components to vary freely instead of constraining $f_s$ to be a scalar. For all lattice depths, we found the superfluid fraction tensor to be nearly isotropic, with $(f_s^{(x,x)}-f_s^{(y,y)})/f_s^{(x,x)}$ and $f_s^{(x,y)}/f_s^{(x,x)}$ typically both below 5$\%$.

\vskip5pt
\paragraph{\textcolor{red1}{Superfluid fraction: dynamic approach--}}
We now describe the determination of the superfluid fraction using an independent transport measurement. In the hydrodynamic regime, in the basis where the superfluid fraction is diagonal, its coefficients are given by \cite{pitaevskiiBoseEinstein2016,hofmann2021hydrodynamics,plattSound2024,poliExcitations2024}
\begin{equation}
    f_s^{(i,i)} = \kappa Mc_i^2,
    \label{eq:fs_hydro}
\end{equation}
where $\kappa = (\rho_0 \partial_{{\rho}_0} \mu)^{-1}$ is the compressibility, a scalar quantity, and $c_i$ is the speed of sound propagating along eigenaxis $i$. In the absence of the lattice, the compressibility is $\kappa_0 = 1/g\rho_0$ and the speed of sound $c_0 = \sqrt{g\rho_0/M}=2.07(5)\,$mm/s, yielding $f_s=1$ as expected for a Galilean-invariant superfluid at $T=0$. The approach used in previous works for 1D modulation \cite{chauveauSuperfluid2023,taoObservation2023}, which relies on taking the ratio of speeds of sound to compute $f_s$, is not applicable to 2D modulations. Therefore, it is necessary to measure the compressibility directly in order to determine the superfluid fraction.

\begin{figure}
    \begin{center}
    \hskip-20pt
        \includegraphics[width=1.05\columnwidth]{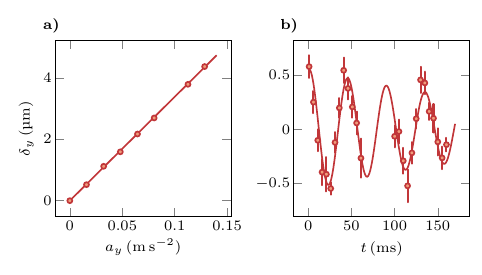}

    \end{center}
    \caption{
    \textbf{Compressibility and speed of sound measurements.} (a) Compressibility determined by measuring the displacement $\delta_y$ of the cloud CoM as a function of the applied static force $M a_y$ to the cloud. The solid line is a fit to the data, giving a slope of $\delta_y / a_y = \SI{34.0 +- 0.7}{\times 10^{-6}\, \s ^2}$. From this slope the compressibility of the cloud is found using Eq.\,\eqref{eq:kappa_meas}. (b) Speed of sound  determined from the oscillation of the CoM, $\delta_y$, after an abrupt release of a static force. The oscillation data is fitted to an exponentially damped sine yielding an oscillation frequency $\nu_y = \SI{22.4 +- 0.2}{Hz}$.
    In both figures, $V_0\simeq 2.4 \,\mu_0$. Injecting the measured slope and oscillation frequency in Eq.\,\eqref{eq:fs_slope+frequ}, we find $f_s^{(y, y)} = \num{0.82 +- 0.02}$.
} 
    \label{fig:delta}
\end{figure}

We measure the compressibility by applying a small constant and uniform force $F_y=M a_y$ along direction $y$ and by extracting the associated Center-of-Mass (CoM) displacement $\delta_y$\,\cite{busleyCompressibility2022}. By combining a local density approximation with a coarse graining over the lattice period, the compressibility is given by \cite{Supplementary} 
\begin{equation}
    \kappa  = \frac{12 \delta_y }{F_y L^2}. 
    \label{eq:kappa_meas}
\end{equation}
Here, the force $F_y$ is produced by applying a static magnetic gradient so that $a_y$ ranges from $0$ to $0.13\,$m.s$^{-2}$. An example of the measurement of $\delta_y$ as a function of $a_y$ for $V_0 \simeq 2.4\,\mu_0$ is reported in Fig.\,\ref{fig:delta}a along with a linear fit to the data. The complete set of measurements of the compressibility as a function of the lattice depth is shown in the End Matter.

We determine the speed of sound by measuring the CoM oscillation frequency in the box potential\,\cite{chauveauSuperfluid2023}. As for the compressibility measurement, we apply a linear magnetic potential along the $y$ direction during the preparation of the cloud, except that it is now abruptly switched off at time $t=0$. We then measure the CoM oscillations of the cloud and obtain its frequency $\nu_y$. Since we mostly excite the fundamental mode of the box, we can deduce the speed of sound from the CoM oscillation frequency, $c_y = 2L\nu_y$. The initial applied force, corresponding to $a_y\simeq 0.019\,$m/s$^{2}$, is significantly smaller than for the compressibility measurement in order to minimize the damping of the oscillations, especially when the lattice depth is high. An example of CoM oscillation for $V_0 \simeq 2.4\,\mu_0$ is reported in Fig.\,\ref{fig:delta}b along with a fit to the data. The complete set of speed of sound  measurements  for several $V_0$ is reported in the End Matter.

Combining the measurement of the compressibility and the speed of sound, we compute the superfluid fraction using Eq.\,\eqref{eq:fs_hydro}, which gives \begin{equation} f_s^{(y,y)}=48\frac{\delta_y} {a_y} \nu_y^2, \label{eq:fs_slope+frequ} \end{equation} where $\delta_y$ is measured when a static force $M a_y$ is applied. The resulting superfluid fraction is reported in Fig.\,\ref{fig:superfluid_fraction} (circles) and is in excellent agreement with the one determined using the previous approach for low lattice depths. Small deviations are observed at large lattice depths. They may be linked to the observed damping of sound waves in this range of parameters (see End Matter), possibly originating from finite temperature effects, the study of which is beyond the scope of this Letter. Both determinations of $f_s$ also lie, within experimental errors, in the region allowed by the measured Leggett bounds.

\vskip5pt
\paragraph{\textcolor{red1}{Conclusion and perspectives--}}
Our work demonstrates the implementation of two methods for determining the superfluid fraction tensor in a 2D modulated weakly-interacting Bose gas. These methods, based on either a dynamic approach or on the direct analysis of the \textit{in situ} density profile of the cloud, are  in good agreement with each other in our experiment. Our approach is not restricted to determining scalar superfluid fractions and can easily be extended to arbitrary lattice potentials. Dipolar supersolids are also good candidates for applying these methods,  for which an approximate  mean-field description is available and their hydrodynamics is well-known\,\cite{blakieSuperfluid2024,poliExcitations2024}. In addition, the method based on the density profile measurements, which is exact for mean-field systems, generally provides an upper bound. It can thus be used  to characterize  strongly interacting Fermi gases, for example, and this bound is tighter than Leggett's one, as discussed for helium supersolids\,\cite{Saslow12}.

\vskip10pt

\begin{acknowledgments}
\textit{Acknowledgments.} 
 We acknowledge the support by ERC (Grant Agreement No 863880) and by ANR (ANR-23-PETQ-0002). We thank Sandro Stringari, Kevin Geier and Thierry Giamarchi for fruitful discussions, Sarah Philips for her participation at the final stage of this project and Jonathan Menssen for critical reading of the manuscript.
\end{acknowledgments}

%% file: End_matter_v9.tex
\subsection{Compressibility measurement}
\begin{figure}[h!!]
    \begin{center}
                \includegraphics[width=0.9\columnwidth]{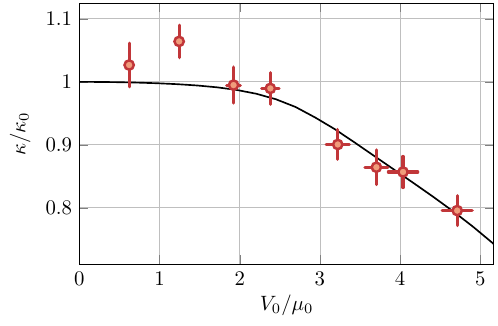}
    \end{center}
    \caption{\textbf{Compressibility measurement.} Normalized compressibility as a function of the normalized lattice amplitude. The solid line represents the corresponding prediction given by the GPE. Error bars correspond to the 1-$\sigma$ statistical uncertainty obtained from typically 40 repetitions of the experiment.}\label{fig:compressibility}
\end{figure}

We show in Fig.\,\ref{fig:compressibility} the measured compressibility normalized to the compressibility $\kappa_0$ of a uniform system. We obtain an excellent agreement with the GPE prediction.

\subsection{Measurement of the CoM oscillation frequency}
\begin{figure}[h!!]
    \begin{center}
        \includegraphics[width=\columnwidth]{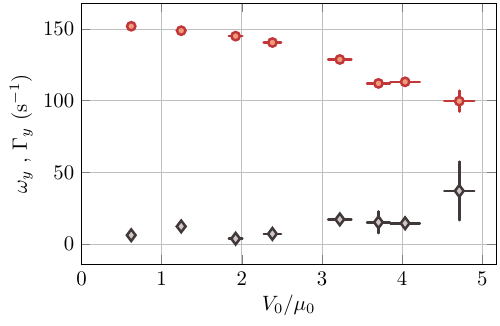}
    \end{center}
    \caption{\textbf{Angular frequency and damping of the oscillations.} Angular frequency (red circles) and damping coefficient (gray diamonds) of the CoM oscillations used to measure the speed of sound. Error bars correspond to the 1-$\sigma$ statistical uncertainty obtained from 10 repetitions of the experiment.
}
    \label{fig:damping}
\end{figure}

\begin{figure}[h!!]
    \begin{center}
        \includegraphics[width=0.95\columnwidth]{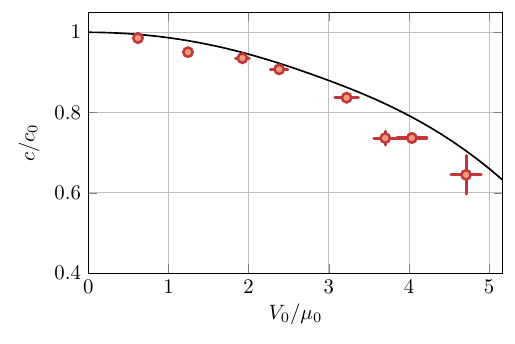}
    \end{center}
    \caption{\textbf{Speed of sound measurement.} Normalized speed of sound as a function of the normalized lattice
    amplitude. The solid line represents the prediction
    given by the GPE. Error bars correspond to the 1-$\sigma$ statistical uncertainty obtained from 10 repetitions of the experiment.}\label{fig:speed_of_sound}
\end{figure}

To extract the oscillation frequency of the cloud confined in the box potential after an excitation along the $y$ direction, we model the time evolution of the CoM of the cloud as a damped free harmonic oscillator\,\cite{villeSound2018,chauveauSuperfluid2023}:
\begin{equation}
\ddot \delta_y+\Gamma_y \dot \delta_y+\omega_y^2 \delta_y=0,
\end{equation}
whose solution for zero initial velocity is 
\vskip-15pt
\begin{equation}
    \delta_y(t) = \delta_{y,0}  +  A \e^{-\Gamma_y t/2} \left[ \cos(\Omega_y t)+\frac{\Gamma_y}{2\Omega_y}\sin(\Omega_y t)     \right],
    \label{eq:fitting_function_com}
\end{equation}
where $\Omega_y=\sqrt{\omega_y^2-\Gamma_y^2/4} $ and assuming $\Gamma_y < 2\omega_y$. We fit our data to this model with $\delta_{y,0}$, $\Gamma_y$, $A$, and $\omega_y$ as fit parameters. The oscillator frequency $\omega_y=2\pi\nu_y$ is used to compute the speed of sound and the superfluid fraction. We show in Fig.\,\ref{fig:damping} the fitted values of $\omega_y$ and $\Gamma_y$ for different lattice depths. Using these results, we determine the speed of sound normalized to the speed of sound in the absence of a lattice, $c_0$, as shown in  Fig.\,\ref{fig:speed_of_sound}. The experimental data is in good agreement with the simulation. At large lattice depths, we notice a deviation to lower values of the measured speed of sound compared to the GPE prediction. This may be due to minor deviations of the lattice potential compared to Eq.\,\eqref{eq:perfect_triangular} or to finite-temperature effects, the study of which is beyond the scope of this Letter. 

\subsection{Upper bounds for the superfluid fraction}

We detail in this paragraph the derivation of the method used in the main text to compute the superfluid fraction for systems at zero temperature and its applicability beyond weakly interacting Bose gases. 

We consider a two-dimensional system of area $L_x\times L_y$ with periodic boundary conditions (PBC) and consisting of $N$ particles. We assume the system has time-reversal symmetry and that its ground state $\ket{\Psi_{\rm eq}}$ does not break this symmetry. The wave function $\Psi_{\rm eq}$ can thus be chosen as real. To compute the superfluid fraction tensor we look for the ground state with twisted boundary conditions (TBC), 
\begin{equation}
\begin{split}
\Psi(\cdots, \boldsymbol{r}_k+ L_x\boldsymbol{e}_x, \cdots) &= \Psi(\cdots, \boldsymbol{r}_k, \cdots)\exp\left(-\mi u_x L_x\right)\\
\Psi(\cdots, \boldsymbol{r}_k+ L_y\boldsymbol{e}_y, \cdots) &= \Psi(\cdots, \boldsymbol{r}_k, \cdots)\exp\left(-\mi u_y L_y\right)
\end{split}
\end{equation}
with $u_xL_x,u_yL_y\ll 2\pi$. 
The superfluid fraction tensor is then given by the energy increase of the ground state at lowest order in $\boldsymbol{u}= u_x \boldsymbol{e}_x + u_y \boldsymbol{e}_y$ \cite{leggettCan1970}, 
\begin{equation}
   \Delta E(\boldsymbol{u})\equiv  E(\boldsymbol{u}) -E_0 = \frac{N\hbar^2}{2M} \sum_{\alpha,\,\beta = x,y}  f_s^{(\alpha,\,\beta)} u_\alpha u_\beta.
    \label{eq:fs_definition_twist}
\end{equation}

To obtain upper bounds on the eigenvalues of the positive symmetric tensor $f_s$, we adopt a variational method. We take the  variational ansatz:
\begin{equation}
    \Psi_{\rm trial}(\boldsymbol{r}_1,\cdots, \boldsymbol{r}_N) = \Psi_{\rm eq} (\boldsymbol{r}_1,\cdots, \boldsymbol{r}_N)\exp\left(\mi \sum_{i=1}^N \tilde S(\boldsymbol{r}_i)\right)
    \label{eq:saslow_var_ansatz}
\end{equation}
where $\tilde S$ is chosen so that $\Psi_{\rm trial}$ satisfies the TBC:
\begin{equation}
    \tilde S(x+L_x,y)=\tilde S(x,y)-u_xL_x, \quad \tilde S(x,y+L_y)=\tilde S(x,y)-u_yL_y.
    \label{eq:TBC_for_S}
\end{equation}
The corresponding energy increase is then
\begin{equation}
     \Delta E_{\rm trial}(\boldsymbol{u})\equiv E_{\text{trial}}(\boldsymbol{u})  - E_0 = \frac{\hbar^2}{2M}\int \rho_{\rm eq}(\boldsymbol{r}) \left(\grad{\tilde S}(\boldsymbol{r})\right)^2\,\dd^2r,
    \label{eq:nbody_trial_energy_simp}
\end{equation}
where $\rho_{\rm eq}(\boldsymbol{r}) = N\int \Psi_{\rm eq}^2(\boldsymbol{r}, \boldsymbol{r}_2, \dots, \boldsymbol{r}_N)\,\dd^2 r_2\dots\dd^2 r_N$ is the particle density. For any choice of $\tilde S$ fulfilling the TBC, $\Delta E_{\rm trial}(\boldsymbol{u})$ gives an upper bound on $\Delta E(\boldsymbol{u})$. The tightest bound is obtained by choosing the function $\tilde S$ that minimizes the trial energy, which obeys
\begin{equation}
    \divergence \left(\rho_{\rm eq}(\boldsymbol{r})\gradient\tilde S(\boldsymbol{r})\right) = 0.
    \label{eq:equation_zero_current}
\end{equation}
This equation can be solved with the constraint given in Eq.\,\eqref{eq:TBC_for_S}, yielding an expression for $\tilde S$ at first order in $\boldsymbol{u}$. The trial energy increase, Eq.\,\eqref{eq:nbody_trial_energy_simp}, is thus of order 2 in $\boldsymbol{u}$, which allows us to define the positive and symmetric rank-2 tensor $g_s$: 
\begin{equation}
\Delta E_{\rm trial}(\boldsymbol{u}) =\frac{N\hbar^2}{2M} \sum_{\alpha,\,\beta = x,y}  g_s^{(\alpha,\,\beta)} u_\alpha u_\beta.
\end{equation}
We introduce the two eigenvalues $f_{s,1},f_{s,2}$ (resp. $g_{s,1},g_{s,2}$) of the tensor $f_s$ (resp $g_s$). Assuming, without loss of generality, $f_{s,1} \leq f_{s,2}$ and $g_{s,1} \leq g_{s,2}$, one can deduce from the general inequality $\Delta E(\boldsymbol{u}) \leq \Delta E_{\rm trial}(\boldsymbol{u})$ for all $ \boldsymbol{u}$ that
\begin{equation}
f_{s,1} \leq g_{s,1} \qquad  f_{s,2} \leq g_{s,2}.
\end{equation}
These inequalities hold even if the eigenvectors of $f_s$ and $g_s$ do not coincide and are valid for any system at zero temperature with time-reversal symmetry.

It is interesting to compare this approach to Leggett's one \cite{leggettCan1970}, which is most relevant when the eigenaxes of $f_s$ are already known. Consider a phase twist along one of these axes, say $x$, so that the total current along the perpendicular axis is zero:
\begin{equation}
    J_y=\frac{\hbar}{M}\int \rho_{\rm eq}(\boldsymbol{r})\,\partial_y \tilde S\,\dd^2 r=0\ .
\end{equation}
The approach above deals with an ansatz for $\tilde S$ that depends both on $x$ and $y$, allowing local currents with arbitrary directions while fulfilling $J_y=0$, whereas the Leggett approach uses a more restrictive ansatz  by assuming that the phase $\tilde S$ depends only on $x$. 

For systems where the ground state satisfies the right-hand side of Eq.\,\eqref{eq:saslow_var_ansatz} at lowest order in $\boldsymbol{u}$ when subjected to twisted boundary conditions, the tensors $f_s$ and $g_s$ coincide. In the case of a weakly interacting Bose gas at zero temperature,  the many-body wave function is given by a product of identical single-particle wave functions obeying the GPE. In the presence of TBC, the wave function can be written as $\sqrt{\rho_{\rm eq}(\boldsymbol r)+\delta \rho(\boldsymbol r)}\e^{\mi \tilde S(\boldsymbol r)}$. The corresponding energy increase is given by the same term as in the right-hand side of Eq.\,\eqref{eq:nbody_trial_energy_simp}, $\hbar^2/(2M)\int \rho_{\rm eq}(\boldsymbol{r}) \left(\grad{\tilde S}(\boldsymbol{r})\right)^2\,\dd^2r$, and a term  $\propto \int(\grad \delta \rho)^2$. The minimum of this second term is obtained independently of $S(\boldsymbol r)$ and for $\delta \rho(\boldsymbol{r})=0$. It justifies that, in this case, $g_s$ coincides with the superfluid fraction tensor $f_s$. We recover the expressions used in the main text by introducing  the phase $S(\boldsymbol{r}) = \tilde S(\boldsymbol{r}) + \boldsymbol{u}\cdot\boldsymbol{r}$, which obeys periodic boundary conditions. Equation \eqref{eq:equation_zero_current} then reads
\begin{equation}
    \divergence \left\{\rho_{\rm eq}(\boldsymbol{r})\left[\gradient S(\boldsymbol{r}) - \boldsymbol{u}\right]\right\} = 0.
    \label{eq:equation_zero_current_shift}
\end{equation}
Integrating by parts in Eq.\,(\ref{eq:nbody_trial_energy_simp}), we find \cite{josserandPatterns2007,josserandCoexistence2007, sepulvedaNonclassical2008,sepulvedaSuperfluid2010,blakieSuperfluid2024}
\begin{equation}
    \Delta E(\boldsymbol{u}) =  \frac{N\hbar^2}{2M} \boldsymbol{u}^2  - \frac{\hbar^2}{2M}\int \rho_{\rm eq}(\boldsymbol{r})\,\boldsymbol{u}\vdot\grad{S(\boldsymbol{r})}\dd^2{r}.
    \label{eq:e_trial_final}
\end{equation}
from which we deduce Eq.\,\eqref{eq:fs_definition_phi}.
\vskip5pt
\paragraph{Isotropy of the superfluid tensor for the triangular lattice.}

For the triangular lattice, the energy $\Delta E(\boldsymbol{u})$, which is a quadratic form in $\boldsymbol{u}$, is invariant under the transformation
\begin{equation}
u_x \to -\frac{1}{2}u_x\pm \frac{\sqrt 3}{2}u_y\qquad u_y \to \mp\frac{\sqrt 3}{2}u_x- \frac{1}{2}u_y\ .
\end{equation}
One immediately deduces that $f_{s}^{(x,x)}=f_{s}^{(y,y)}$ and $f_{s}^{(x,y)}=0$.

%% file: sm_v13.tex
\subsection{Determination of the compressibility}

We detail in this paragraph the determination of Eq.\,\eqref{eq:kappa_meas} of the main text. In the presence of a periodic potential $V(\boldsymbol{r})$, the GPE for the many-body wave function $\psi(\boldsymbol{r})$ reads
\begin{equation}
    -\frac{\hbar^2}{2M} \bnabla^2 \psi(\boldsymbol{r}) + g|\psi(\boldsymbol{r})|^2 \psi(\boldsymbol{r}) + V(\boldsymbol{r}) \psi(\boldsymbol{r})= \mu \psi(\boldsymbol{r}),
    \label{eq:GPE}
\end{equation}
where $\mu$ is the chemical potential and $\int |\psi(\boldsymbol{r})|^2\mathrm{d}\boldsymbol{r}=N$ for a given atom number $N$. When an additional force $F_y$ is applied along the $y$ direction, the previous equation is modified by adding a term $-F_y y\psi(\boldsymbol{r})$ to the right-hand side. If, over a unit cell of the potential, the variation of $F_y y$ is much smaller than $\mu$, one can write
\begin{equation}
    -\frac{\hbar^2}{2M}\bnabla^2 \psi(\boldsymbol{r}) + g|\psi(\boldsymbol{r})|^2 \psi(\boldsymbol{r}) + V(\boldsymbol{r}) \psi(\boldsymbol{r})= \left(\mu-F_y y_c\right) \psi(\boldsymbol{r}),
    \label{eq:GPE_aprox_UC}
\end{equation}
where $y_c$ is the cell center. Each unit cell thus has a local chemical potential $\mu-F_y y_c$. The mean density over the cell is given by 
\begin{equation}
    \left< \rho \right>_{\text{UC}}
    =\rho_0 - \left.\frac{\partial \rho}{\partial \mu}\right|_{\mu} F_y y_c
    = \rho_0 \left( 1 - \kappa F_y y_c \right),
    \label{eq:n_LDA}
\end{equation}
where $\rho_0$ is the average density over the whole system and $\kappa$ its compressibility. The CoM displacement induced by the force $F_y$ is given by 
\begin{equation}
    \delta_y = \frac{1}{N}\iint \mathrm{d}x\mathrm{d}y  \, y\,[\rho(x,y)-\rho_0].
    \label{eq:eq_com_shift}
\end{equation}
We also assume that the system is symmetric with respect to the origin and is periodic with period $d$ along $y$. Splitting the integration over each unit cell and using Eq.\,\eqref{eq:n_LDA}, we obtain
\begin{equation}
    \delta_y = \frac{\rho_0\kappa F_y L }{N} d\sum_{i=-L/2d}^{L/2d} (i d)^2 \simeq \frac{\kappa F_y L^2}{12} .
    \label{eq:eq_com_shift_kappa}
\end{equation}

\subsection{Optical lattice potential generation}
The characteristics of the optical system can be found in the Supplemental Material of Ref.\,\cite{chauveauSuperfluid2023}. The optical lattice potential imprinted on the atoms is approximately given by Eq.\,\eqref{eq:perfect_triangular} in the main text. To imprint this potential, we display the following profile on the DMD, which acts as an amplitude modulator \cite{dorrerDesign2007}:
\begin{equation}
    f(\boldsymbol{r}) = s(\boldsymbol{r})\sqrt{B - A\sum_{m=0}^{2} \cos(\boldsymbol{k}_m \cdot \boldsymbol{r}+\phi_m)},
    \label{eq:DMD_function}
\end{equation}
where $A$ and $B$ are positive coefficients chosen such that $\forall \boldsymbol{r}$, $f^2(\boldsymbol{r})\in [0.1 , 1]$. A proper choice of the spatial origin allows us to set $\phi_1=\phi_2=0$. We also choose $\Phi=0$, but imperfections in the optical system imaging the DMD pattern onto the atoms lead to a modified potential with a nonzero value of $\Phi$, calibrated as shown in the next section. The function $s(\boldsymbol{r})$ is determined in a separate calibration experiment to make the laser beam profile reflected off the DMD uniform over the size of the cloud. Since the DMD is a binary amplitude modulator, we obtain a smoothly varying profile following Eq.\,\eqref{eq:DMD_function} using a dithering method~\cite{dorrerDesign2007}. The coefficients $A$ and $B$ and the correction function $s(\boldsymbol{r})$ remained constant during data acquisition; the lattice amplitude was tuned by adjusting the power of the light illuminating the DMD. 

\subsection{Optical lattice potential characterization}
We use an auxiliary CCD camera, which provides an image of the trapping light profile in the atomic plane, to monitor and calibrate the potential imposed on the cloud. We first compare the measured intensity profile to the one introduced in Eq.\,\eqref{eq:perfect_triangular} of the main text. We fit the data to the following function:
\begin{equation}
\begin{split}
    I(\boldsymbol{r})=\bar{I}+I_0 \cos(\boldsymbol{k}_0 \cdot \boldsymbol{r})+I_1 \cos(\boldsymbol{k}_1 \cdot \boldsymbol{r})\\+I_2 \cos(-\boldsymbol{k}_0 \cdot \boldsymbol{r}-\boldsymbol{k}_1 \cdot \boldsymbol{r}+\Phi),
    \label{eq:potential_real}
\end{split}
\end{equation}
where the phases in the first two cosines have been set to zero by a proper choice of the spatial origin. We obtain $I_2/I_0=1.00(2)$, $I_1/I_0=0.98(2)$, $k_1/k_0=0.96(1)$, $\Phi=0.21(1)\pi$, and an angle $\vartheta=0.671(6)\pi$ between the two vectors $\boldsymbol{k}_0$ and $\boldsymbol{k}_1$. This measurement justifies, within error bars, the choice of potential in Eq.\,\eqref{eq:perfect_triangular} of the main text. For all computations with the GPE shown in this work, we consider the isotropic lattice given in Eq.\,(2) of the main text with $\Phi=0.21\pi$.

\subsection{Calibration of the imaging system and reconstruction of the density profile}
As introduced in the main text, the finite imaging resolution of the atomic cloud must be taken into account when we determine the superfluid fraction or Leggett bounds from the measured density profile. We write this profile as
\begin{equation}
\rho^{\text{meas}}(\boldsymbol{r}) = \rho_0 + \sum_{n>0} \beta_n \rho_n (\boldsymbol{r}),
\end{equation}
where the $\beta_n$ are the attenuation factors associated with a given spatial frequency with index $n$ of the density modulation. We calibrated the $\beta_n$ using the same method as described in Ref.\,\cite{chauveauSuperfluid2023}, which we briefly summarize here.

First, we calibrate the light intensity level using the atomic response to a lattice with a large period of $\SI{24}{\micro \meter}$, for which we can safely assume that the corresponding $\beta$ coefficient is one. We use shallow lattice potentials so that the induced density modulation $\Delta \rho$ is well approximated by
\begin{equation}
\frac{\Delta \rho}{\rho_0}\approx\frac{V_0}{\mu_0+\varepsilon_k/2},
\end{equation}
where $\varepsilon_k=\hbar^2 k^2/2M$ is the recoil energy of the triangular lattice of period $4\pi/(\sqrt{3}k)$. We then compare the atomic response for different lattice periods to this reference measurement and obtain $\beta_1=0.71(4)$ and $\beta_2=0.36(16)$, corresponding to lattice periods of $\lambda_1=d = \SI{6}{\micro \meter}$ and $\lambda_2 \simeq \SI{3.46}{\micro \meter}$, respectively. For smaller periods, the values of the $\beta_n$ are compatible with zero within error bars.

\begin{figure}[t!!]
    \begin{center}
        \includegraphics[width=0.8\columnwidth]{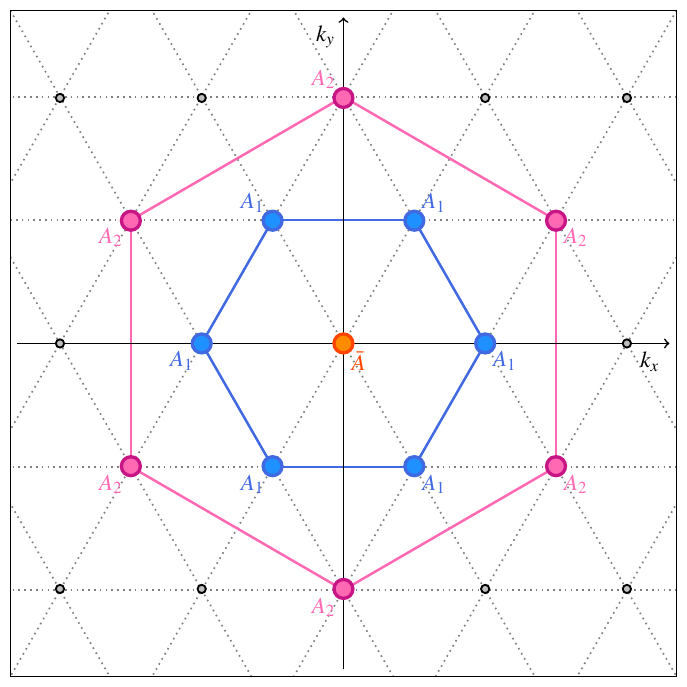}
    \end{center}
    \caption{\textbf{Fourier components used for fitting the density profile.} We fit the density
        profiles using a sum of two triangular lattices of different periods and an average density $\bar A$. The amplitude of the components of the triangular
        lattice of period $\lambda_1=6\,\mum$ (resp.  $\lambda_2\simeq3.46\,\mum$) are all equal to $A_1$ (resp. $A_2$) to ensure the isotropy of the deduced superfluid fraction tensor. The norm and orientation of the wave vectors of the lattices are also fixed according to this figure.
    }
    \label{fig:Fourier}
\end{figure}

We determine the superfluid fraction and Leggett bounds from \textit{in situ} images using a reconstructed density profile that we obtain from a fit of the experimental images to a function including the first two spatial frequencies of the triangular lattice and taking into account the correction by the attenuation factors $\beta_{1,2}$ of the imaging system. The fitting function is given by 
\begin{eqnarray}
n(\boldsymbol{r})=\bar A&+& A_1\sum_{\alpha=0}^{\alpha=2}  \cos[\boldsymbol{k}^{(1)}_{\alpha} \cdot (\boldsymbol{r}-\boldsymbol{r}_0)+\varphi_{1}] \nonumber\\
&+& A_2\sum_{\gamma=0}^{\gamma=2}  \cos[\boldsymbol{k}^{(2)}_{\gamma} \cdot (\boldsymbol{r}-\boldsymbol{r}_0)+\varphi_{2}]
\end{eqnarray}
It corresponds to an average density $\bar A$ modulated by the sum of two triangular lattices with period $\lambda_1=2\pi/ |\boldsymbol{k}^{(1)}_{\alpha} |$ and $\lambda_2=2\pi/ |\boldsymbol{k}^{(2)}_{\alpha}|$. The $\boldsymbol{k}^{(1)}_{\alpha}$ are defined in Eq.\,\eqref{eq:perfect_triangular} of the main text, and we choose $\boldsymbol{k}^{(2)}_{i}=\boldsymbol{k}^{(1)}_{i}-\boldsymbol{k}^{(1)}_{i-1}$ (with circular permutation of the indices).  We ensure that this function is invariant under a rotation by an angle of $2\pi/3$ around $\boldsymbol{r}_0$ by imposing the same amplitude $A_{1,2}$, the same norm $|\boldsymbol{k}_{\alpha,\,\beta}|$, and the expected orientation of the $\boldsymbol{k}_{\alpha,\beta}$ (see Fig.\,\ref{fig:Fourier}) for the three terms of each of the two triangular lattices. The free parameters of the fit are thus $\bar A$, $A_1$, $A_2$, $\boldsymbol{k}_0^{(1)}$, $\boldsymbol{r}_0$, $\varphi_1$ and $\varphi_2$.

For the range of parameters explored in this work, $V_0<5\,\mu_0$ and $f_s>0.4$, we verified numerically that the truncation to the Fourier components show in Fig.\,\ref{fig:Fourier} is sufficient to provide a robust estimate of these quantities.   For large $V_0$, this truncation is expected to slightly underestimate the $f_s^-$ bound, which is the most sensitive to it. The truncation to these first two Fourier components also leads to an unphysical reconstructed density profile with negative values for sufficiently large potential depth. We address this issue by setting points with negative density values to a small arbitrary nonzero value $\epsilon=10^{-25}$ after normalizing the profiles to a maximum value of 1. We confirmed numerically that this procedure gives a stable determination of the studied quantities for any sufficiently small value of $\epsilon$. 

\subsection{Comparison of Leggett bounds to numerical predictions}
\begin{figure}[t!!]
    \begin{center}
        \includegraphics[width=\columnwidth]{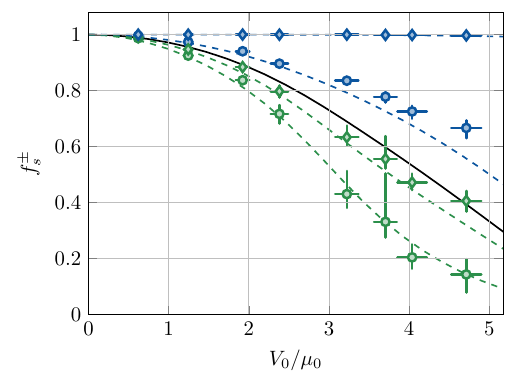}
    \end{center}
    \caption{\textbf{Leggett bounds.} Measured Leggett bounds corresponding to the upper
        bounds  $f_s^{+} (\boldsymbol{e_y})$ (blue diamonds) and $f_s^{+} (\boldsymbol{e_x})$ (blue
        circles), and to the lower bounds $f_s^{-} (\boldsymbol{e_y})$ (green diamonds) and $f_s^{-} (\boldsymbol{e_x})$ (green
        circles).  Error bars correspond to the error propagation of the uncertainties in the calibration of the imaging system response and of the lattice. The dashed lines are the prediction for these bound computed on the profiles obtained from the simulations of the GPE. The solid line is the predicted superfluid fraction.}
    \label{fig:BoundsGPE}
\end{figure}

We show in Fig.\,\ref{fig:BoundsGPE} the Leggett bounds computed from the density profile for $\theta=0,\pi$. These data are in good agreement with the predictions of the GPE (dashed lines). 
